# On the Origin of Metallicity and Stability of the metastable phase in Chemically Exfoliated MoS$_2$

Debasmita Pariari [a], Rahul Mahavir Varma [a], Maya N. Nair [a,+], Patrick Zeller [b], Matteo Amati [b], Luca Gregoratti [b], Karuna Kar Nanda [c], D. D. Sarma [a,*]

[a]*Solid State and Structural Chemistry Unit, Indian Institute of Science, Bengaluru – 560012.*

[b]*Elettra-Sincrotrone Trieste S.C.p.A., SS14, km 163.5 in AREA Science Park, 34149 Basovizza, Trieste, Italy.*

[c]*Materials Research Centre, Indian Institute of Science, Bengaluru – 560012.*

[*]Corresponding author: sarma@iisc.ac.in

[+]Present address: CUNY Advanced Science Research Center, 85 St. Nicholas Terrace, New York, NY 10031, USA

Abstract:

Chemical exfoliation of MoS$_2$ via Li-intercalation route has led to many desirable properties and spectacular applications due to the presence of a metastable state in addition to the stable H phase. However, the nature of the specific metastable phase formed, and its basic charge conduction properties have remained controversial. Using spatially resolved Raman spectroscopy (~1 μm resolution) and photoelectron spectroscopy (~120 nm resolution), we probe such chemically exfoliated MoS$_2$ samples in comparison to a mechanically exfoliated H phase sample and confirm that the dominant metastable state formed by this approach is a distorted T′ state with a small semiconducting gap. Investigating two such samples with different extents of Li residues present, we establish that Li$^+$ ions, not only help to exfoliate MoS$_2$ into few layer samples, but also contribute to enhancing the relative stability of the metastable state as well as dope the system with electrons, giving rise to a lightly doped small bandgap system with the T′ structure, responsible for its spectacular properties.

1. Introduction

Amongst the non-graphene layered materials, transition metal dichalcogenides (TMDs) have been in the focus of the research community for quite a long time because of their fascinating properties and novel applications that emerge upon 2D confinement. Specifically, molybdenum disulphide ($MoS_2$) in this series bears great promises for displaying wide range of electronic [1–5], mechanical [6–9], optical [10–15] and chemical properties [16,17] due to the presence of several polymorphs with distinctly different electronic structures [18–20] including the possibility of exotic properties such as the quantum spin Hall effect [21]. $MoS_2$ structure is built from two close-packed planes of $S^{2-}$ ions, sandwiching the hexagonal arrangement of $Mo^{4+}$ layer. Depending on the relative positioning of the two sulphur planes with respect to each other, two basic polymorphs of $MoS_2$ are realised, with the H phase having the ABA arrangement and the T phase with the ABC arrangement of the $S^{2-}$-$Mo^{4+}$-$S^{2-}$ trilayers. Thus, the coordination of sulphur ions around the central Mo sites are distinctly different with the H phase having a trigonal prismatic coordination and the T phase an octahedral one. While H is the thermodynamically ground state, octahedral T structure can undergo further distortion to minimize the energy, consistent with the simple molecular orbital picture [22]. Such distortions lead to the formation of various superlattices with $1 \times 2$ superlattice with chains of dimers (T′), $2 \times 2$ (T″) and $\sqrt{3} \times \sqrt{3}$ (T‴) periodic units relative to the undistorted $1 \times 1$ (T) unit cell [23,24].

$MoS_2$, showing semiconducting property with a tuneable band gap of ~ 1.2 – 1.9 eV depending on the number of layers present [25,26] in its H phase can undergo phase transition involving different routes, such as, plasma hot electron transfer [27,28], mechanical strain [29–31], electron-beam irradiation [18,19] and chemical reactions, often at the expense of its semiconducting property. The chemical routes normally involve non-aqueous alkali metal intercalation [32–34] followed by a water exfoliation step, sophisticated electrochemical control [15,35] or expansion of the interlayer distance by hydrothermal synthesis [36,37]. Although theoretical calculations predict that different superlattices of the metastable state show different properties such that T′ and T″ are small band gap semiconductor [38,39], T‴ is a ferroelectric insulator [40], it has been generally assumed that, all the above mentioned processes lead to the formation of the metallic T phase [27,41–43]. However, theoretical calculation predicts T-phase to be not a metastable phase, but an intrinsically unstable state with spontaneous distortions lowering its energy [40].

Direct experimental evidence about the electronic structure of the metastable states has been obtained by probing the states close to Fermi level with photoemission spectroscopy [32]. It was conclusively shown in this paper from both experimental and theoretical considerations that, the lithium intercalation in the $MoS_2$ sheets leads to the formation of semiconducting T′ state instead of metallic T state. But lithium intercalated $MoS_2$ has been extensively studied for different application purposes such as hydrogen evolution reaction [44,45] and supercapacitor [46,47] to name a couple, based deeply in the belief of its metallic property. Thus, there is a clear divergence of views along two mutually incompatible scenarios for the metastable phase found in relative abundance in chemically exfoliated $MoS_2$. One of these views is based on its many useful device properties, subscribing to a metallic and therefore, the undistorted T phase. The opposing viewpoint stresses the impossibility of the T phase due to its intrinsic instability, suggesting instead that the metastable phase is in fact a small bandgap, semiconducting T′ phase. The T′ phase is derived from the T phase by a simple dimerization, leading to alternate long and short Mo-Mo distances, as shown in Fig. S1 of the Supplementary Material. In all these discussions, both camps across the dividing line have generally ignored any possible role of $Li^+$ ions that are used in the process of the chemical exfoliation. We focus our attention on this very aspect to establish that $Li^+$ ions play a crucial role in this context and our results provide a natural resolution of the apparently conflicting viewpoints described above in terms of a metallic state achieved via $Li^+$ ion induced doping of electrons into the small bandgap, semiconducting distorted T′ phase.

2. Methodologies

Flakes of the stable H phase $MoS_2$ were obtained by the micromechanical scotch tape exfoliation and were deposited on conducting indium doped tin oxide (ITO) substrates, to avoid the sample from getting charged during photoemission experiments. We characterised several such flakes using AFM and found that most of these have about 4-9 layers of $MoS_2$ with a typical interlayer separation of about 0.65 nm [48,49](see Fig. S2 a-f in Supplementary Material). The substrate, with the flakes on top of it, are then sealed in a vial at inert atmosphere and treated with 1.6$M$ $n$-butyl lithium in hexane solution for two days at room temperature. In this step $Li_xMoS_2$ is formed. Without exposing the samples to the ambient, the unreacted $n$-butyl lithium is then removed from the sample by injecting hexane repeatedly inside the vial and washing it thoroughly. Finally, the water exfoliation step is done by removing the sample from the vial and washed several times with distilled water to remove any excess intercalated $Li^+$ in the form of lithium hydroxide (LiOH).

In order to produce a series of samples having different concentration of $Li^+$, we kept everything same in the above-mentioned process except the last water exfoliation step. We washed the samples with distilled water through different numbers of wash-cycles to remove the extra $Li^+$ from the system, namely 12 times, 6 times, 4 times, 1 time and without any water-wash at all. We refer throughout the text to these samples as 12W, 6W, 4W, 1W and 0W, respectively. In order to ascertain the consequence of $Li^+$ intercalation and subsequent washing with water in effectively separating the layers of $MoS_2$ by increasing the van der Waals gap, we investigated several flakes before intercalation, after intercalation and after the water washing (see Fig. S3 a-f in Supplementary Material). These results show that the interlayer gap increases to typically 1.3-1.5 nm and 0.9-1.1 nm with the intercalation and washing, respectively; this markedly enhanced interlayer separation compared to that (0.65 nm) of the H stable phase ensures electronically decoupled single layers of $MoS_2$ in the chemically treated samples, as reported earlier [48].

The reaction of $Li_xMoS_2$ with water is so vigorous that no trace of $Li^+$ is found in photoemission spectra after washing the sample just once with distilled water (see Fig. S4 in Supplementary Material), leading to very similar results for the 12W, 6W, 4W and 1W samples, as shown in Fig. S5 in Supplementary Material in terms of Raman spectra of all compounds. Even electron spectroscopic results between the 1W and 12W samples are very similar, while being distinct from the 0W sample, as illustrated in Fig. S6 in Supplementary Material. Therefore, we discuss below results obtained only from 0W and 12W samples, where we have collected data with the highest signal-to-noise ratio. For the sake of comparison with a reference sample, we carried out a complete set of experiments with identical conditions on mechanically exfoliated $MoS_2$ flakes without the lithium treatment; this sample is referred as H in this paper.

Micro-Raman measurements were carried out with LabRAM HR Evolution (HORIBA) confocal microscope-based Raman spectrometer at room temperature with a beam spot of radius ~1 μm. All Raman spectra were recorded with the excitation laser wavelength and power of 532 nm and 1.2 mW, respectively. Scanning photoelectron microscopy (SPEM) experiments were performed at the ESCA microscopy beamline of Elettra Synchrotron laboratory, with hν = 520 eV, spatial resolution of ~120 nm and energy resolution of 180 meV. While detailed spectral features were obtained from specific spots on samples, the spot size is limited by the photon beam size (~120 nm). Composition resolved maps and images of the samples were obtained over a much larger dimension, typically about 40 μm x 40 μm area, by restoring the photon beam across the sample face with the help of a piezo-driven sample mounting stage. All measurements were carried out with a chamber pressure ~ $10^{-10}$ mbar at ambient temperature.

3. Results and Discussions

Raman spectroscopy has proved to be a very powerful technique to understand the vibrational fingerprint of 2D layered materials. In Fig. 1a. Raman spectra, normalised at $A_{1g}$ peak, obtained from three different samples: (i) H (blue circles), (ii) 12W (red triangles) and (iii) 0W (black dots), are compared. In order to avoid conversion of the metastable phase to the stable one under the excitation source, it was found necessary to record all data using a very low laser power (1.2 mW), explaining the relatively noisy data; however, the peaks and their positions are unambiguously identifiable in these spectra. H sample exhibits two peaks at 380 and 404 cm$^{-1}$ due to $E_{2g}^1$ and $A_{1g}$ responsible for in plane and out of plane opposite vibrations of S and Mo atoms, respectively. For intercalated samples in the lower wavenumber region, four additional peaks, commonly termed $J_1$ (156 cm$^{-1}$), $J_2$ (220 cm$^{-1}$), $E_{1g}$ (283 cm$^{-1}$) and $J_3$ (330 cm$^{-1}$), are clearly observed. The appearance of these additional Raman peaks only from chemically exfoliated samples in Fig. 1a is a clear evidence of additional polymorphs of MoS$_2$ in such cases compared to the pure H phase of the mechanically exfoliated samples. Such results have been extensively reported in the literature and often these new Raman lines have been used to argue the presence of metallic T phase [45,50]. However, this can be easily seen to be erroneous in the following way. The point group describing the symmetry of a single MoS$_2$ layer of undistorted T-phase is $D_{3d}$. Applying arguments from group theory, irreducible representation for normal modes of vibrational degrees of freedom can be written as:

$$\Gamma_{vib}(1T) = A_{1g} + E_g + 2E_u + 2A_{2u}$$

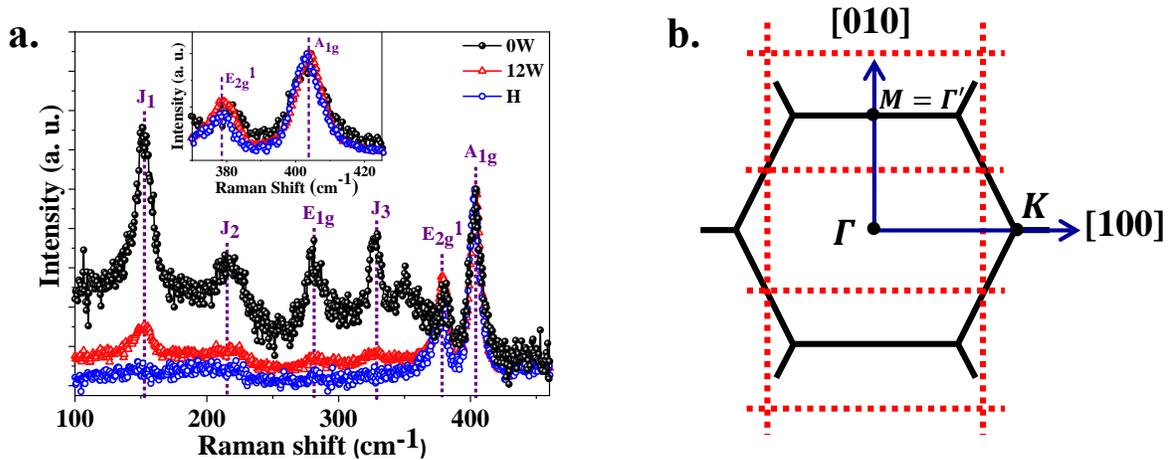

**Fig. 1.** (a) Room temperature Raman Spectra for pristine H and both of the intercalated samples (0W and 12W). The inset shows the spectral features in the region of about 370-425 cm$^{-1}$ in an expanded scale for clarity; (b) Reciprocal space diagram illustrating the relationship between the basal plane projection of the Brillouin zone of H-MoS$_2$ (solid black lines) and the Brillouin zone of a $2a_0 \times a_0$ superlattice (red dotted lines) corresponding to the T′ structure.

By looking at the basis functions of each irreducible representations from the character table of $D_{3d}$ point group, A$_{1g}$ and E$_g$ modes can be easily recognised as Raman active modes, whereas E$_u$ and A$_{2u}$ modes are infra-red active. This analysis is also supported by several theoretical study described in the literature [23]. So, only two peaks should appear in Raman spectra of the undistorted T-phase MoS$_2$. In Fig. 1a, the presence of four peaks, namely, J$_1$, J$_2$, E$_{1g}$ and J$_3$ peaks, clearly indicates that more than two vibrational normal modes are Raman active for the metastable state present in the chemically exfoliated samples. This is a strong evidence that the metastable state present in this case cannot be the undistorted T-phase. In fact, the frequencies of these four additional peaks in the Raman spectra of chemically exfoliated samples suggests the formation of the distorted T′ phase.

The appearance of the additional peaks can be explained in terms of the presence of the superlattice in such distorted structures. Because of wave vector conservation, the peaks in Raman spectra are normally responsible for the normal modes vibrations at the zone centre (k = 0) [51]. But if a superlattice is formed, the Brillouin zone is reduced, and zone boundary points of the undistorted lattice can coincide with the centre of the new zone. A typical example is shown in Fig. 1b, for the formation of 2 × 1 superlattice of the T′ phase. In this case, the M point of the original Brillouin zone is folded to the centre of the Brillouin zone of the distorted crystal. As a result, apart from $\Gamma$ point, vibrational modes from M can also emerge as Raman peaks. For the T′-phase irreducible representation for the vibrational normal modes can be written [52],

$$\Gamma_{vib}(1T') = 6A_g + 3B_g + 3A_u + 6B_u$$

where, $A_g$ and $B_g$ modes are Raman active. Therefore, group theory predicts the appearance of as many as nine peaks for the T′-$MoS_2$ sample, which is in good agreement with the theoretical calculations [23]. Within the interpretation of the formation of primarily T′ phase as the metastable polymorph in chemically exfoliated $MoS_2$, $J_1$ peak is composed of two different phonon modes, namely out of plane opposite vibration of each stripe of the Mo atoms inside the zigzag chain and in plane shearing vibration of one stripe of atoms with respect to another inside a chain. $J_2$ peak is responsible for the phonon modes which tends to restore the original trigonal prismatic structure of H phase by equalising Mo - Mo distance between two zigzag chains. Though theoretically predicted [39] most intense peak is $J_2$, the much larger full width at half maxima (FWHM) has substantially reduced the height of the peak. $E_{1g}$ peak arises from the anti-symmetric stretching of the adjacent S atom layers with respect to the Mo atom inside a zigzag chain. The slight out of plane phonon mode responsible for $J_3$ peak tends to break each zigzag chain from its adjacent one in two stripes.

It is interesting to note here the intense presence of the additional Raman peaks in the 0W sample and its significant reduction in the 12W sample. Since the only difference between these two samples is washing of the 12W sample with distilled water at room temperature in order to remove $Li^+$ ions, it would appear that the presence of $Li^+$ ions is essential to give rise to the extensive presence of the metastable state, evidenced by the strong presence of the characteristic, four additional Raman peaks (Fig. 1a) and the removal of the $Li^+$ ions, as in 12W sample, leads to a sizable conversion of the metastable state to the stable H phase, as evidenced by the reduction of these Raman signals. We note that this is not a thermally-induced gradual conversion of the metastable polymorph to the stable one merely as a function of time, as explicitly checked by us by recording Raman spectra of the 0W sample after the lapse of a period of time equivalent to the time required to obtain 12W sample from a 0W one. Moreover, we also note that the Raman spectra of the 1W sample also show a similar reduction of the characteristic Raman signals. Thus, the presence of $Li^+$ ions, not only facilitates reaching the metastable phase kinetically, but it also enhances its relative stability thermodynamically. We shall discuss this point further after presenting electron spectroscopic evidences in the following parts.

Photoelectron spectra in the Mo 3*d* region for the two chemically exfoliated samples, namely 0W (black dots) and 12W (red triangles), are presented in Fig. 2a along with the spectrum (blue circles) from the mechanically exfoliated H sample using the highly focused photon beam from an arbitrary point on each sample; it was confirmed that spectral features remain qualitatively the same at different points on several flakes of the same sample. We can easily image the particular flake under investigation by mapping the integrated intensity of the Mo $3d_{5/2}$ peak between 226.84 and 230.7 eV over the entire sample area; the resulting Mo 3*d* intensity maps for flakes of 0W and 12W samples, most investigated for the purpose of this work, are shown in Fig. 2b and 2c. These images clearly show the well-defined presence of the two flakes on the ITO substrate, illustrated by the approximate boundary drawn with the cyan dotted lines.

The reference H sample exhibits two narrow peaks at binding energies of 229.4 eV (Mo $3d_{5/2}$) and 232.5 eV (Mo $3d_{3/2}$) and a broad, low-intensity feature at ∼226.6 eV due to S 2*s* contributions (Fig.

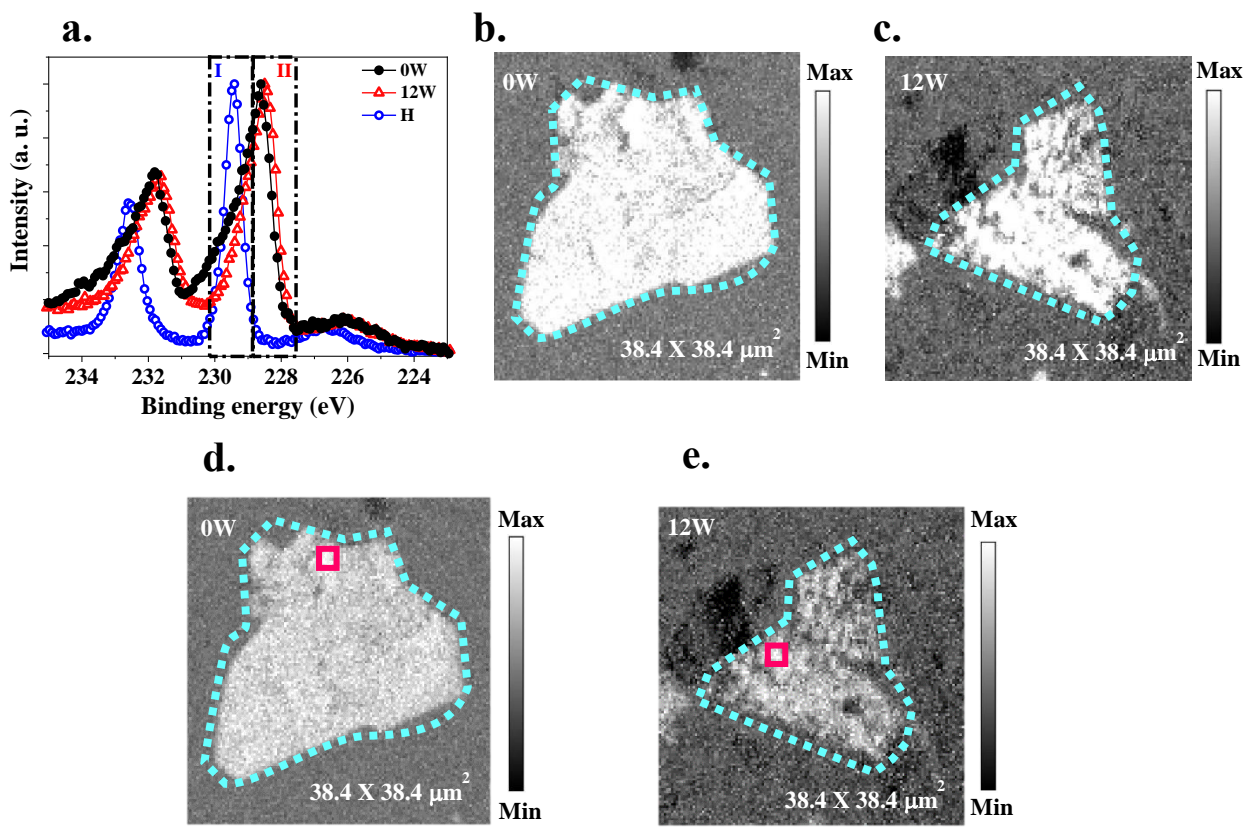

**Fig. 2.** (a) Mo 3*d* core level spectra for 0W, 12W and H sample; Map of Mo $3d_{5/2}$ signal intensity over the binding energy range of 226.84-230.7 eV in the (b) 0W sample and (c) 12W sample; Map of the intensity ratio II/I over respective energy windows shown in (a) for (d) 0W and (e) 12W sample.

2a). Upon comparing this spectrum with the spectra taken on the chemically treated samples, one observes two major differences: (i) spectra from the intercalated samples are prominently asymmetric and shifted towards the lower binding energy, (ii) FWHM of these spectra are large compared to the spectrum obtained from the pristine reference H-phase sample. These are clear indications that in case of intercalated samples, the dominant species is chemically shifted with respect to the reference sample, indicating an extensive presence of a metastable polymorph. Larger widths of the spectral features suggest the presence of multiple phases with finite chemical shifts between signals arising from different polymorphs. This is consistent with the observation described in ref. 32, where two distinct phases of MoS$_2$, namely H and T′, were identified as contributing to the spectral features of the chemically exfoliated MoS$_2$ in contrast to the almost exclusive presence of H- phase signal in mechanically exfoliated MoS$_2$ [53]. Following the same procedure, Mo 3$d_{5/2}$ peak for the intercalated samples in Fig. 2a can be divided in two different energy windows based on dominant contributions of the two different phases. It is clear that the intensity in the region I (228.9 eV – 230.1 eV) is dominated by contributions from the H-phase whereas contribution in region II (227.5 eV - 228.9 eV) is mainly from the metastable T′-phase with hardly any contribution from the stable H phase, as indicated by the spectral features of the pure H phase sample in Fig. 2a. Therefore, the contrast in the intensity ratio II/I, plotted in Fig. 2d and 2e for 0W and 12W, respectively, reveals the relative abundance of the metastable phase in the intercalated samples, with the lighter regions corresponding to the T′ MoS$_2$-rich regions. Fig. 2d shows that the T′-phase is reasonably uniformly distributed over the entire 0W sample in abundance. In contrast, Fig. 2e shows an overall markedly reduced abundance of the T′-phase in the 12W sample with some parts of the sample showing even more pronounced absence of the metastable phase than other parts, leading to a patchy distribution of the metastable phase. This is consistent with the conclusions drawn from the Raman studies with the 12W sample showing a significantly reduced average metastable phase contributions. This suggests that washing away of $Li^+$ ions with water indeed helps to convert metastable T′-phase into the stable H-phase or conversely, retaining $Li^+$ ions helps to stabilise the metastable state.

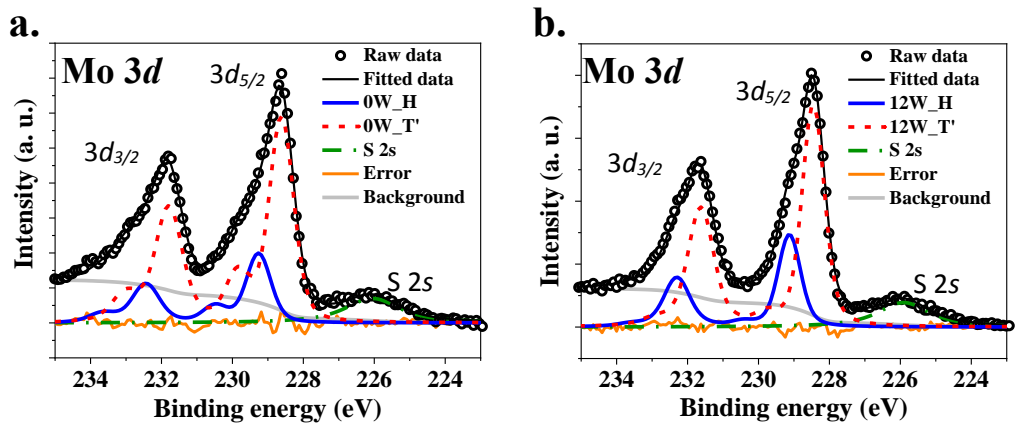

**Fig. 3.** Mo 3$d$ core level spectra (black open circle) and fitting of these spectra with two contributions for (a) 0W and (b) 12W samples. S 2$s$ contribution is also shown.

In order to understand the nature and the origin of the metastable T′ phase, we identify the photoemission pixel with the largest T′ contributions in both samples, enclosed by the solid rectangle, shown in Fig. 2d and 2e and analyse spectra obtained from these pixels in detail. The Mo $3d$ spectra obtained from the specified spots are shown in Fig. 3a and 3b for 0W and 12W samples, respectively; we note that these spectra are the same ones shown in Fig. 2a overlapping with each other for an easy comparison. Our initial attempt to describe the Mo $3d$ spectral region as a sum of two chemically-shifted spectral contributions arising from the stable H and the metastable T′ phases did not work at all for the spectral features from the 0W sample, indicating the presence of additional spectral features in this case. In fact, a closer inspection of Fig. 2a, that overlaps the two spectra in Fig. 3a and 3b, reveals additional spectral weights for the 0W sample compared to the 12W one on the higher binding energy side, for example around 230 eV for the Mo $3d_{5/2}$ and around 233 eV for the Mo $3d_{3/2}$ regions. A reliable description of the spectral features over the entire energy therefore requires additional components to account for such high binding energy satellite features accompanying the main peaks. Since spectral analysis of the S $2p$ region, discussed later in this paper, also requires similar satellite features with similar intensity and peak position relative to the main peak as in the Mo $3d$ region, it appears that there are prominent plasmon loss features accompanying all photoelectron signals from the 0W sample. Therefore, we fitted the spectral features obtained from 0W sample in Fig. 3a in terms of contributions from two phases, H and T′, each consisting of S $2s$, Mo $3d_{5/2}$ and Mo $3d_{3/2}$ main peaks with each main peak accompanied by plasmon loss satellite replicas with a fixed relative intensity and peak position; details of the spectral fitting have been described in greater detail in the SI. Allowing the relative intensity and the peak position of the plasmon satellite to vary, we obtain a very good description of the entire spectrum as shown in Fig. 3a with 0.23 and 1.3 eV as the relative intensity and the peak position for the plasmon satellite relative to the corresponding main peak, with the fit and the components also shown in Fig. 3a. An identical analysis of the spectrum recorded from the 12W sample yields an equally good fit of the entire spectrum with a relative intensity and peak position of 0.05 and 1.2 eV for the plasmon satellite, as shown in terms of the best fit and corresponding component spectra in Fig. 3b. Thus, there appears to be a considerable decrease in the plasmon satellite contribution in the 12W sample compared to the 0W sample. The lower intensity of the plasmon satellite in the 12W sample compared to the 0W sample suggests that the electron density responsible for this plasmonic signature is in higher concentration for the 0W sample compared to the 12W sample. In fact, we find that the spectral features for the 12W sample can be described reasonably well even without invoking any plasmonic satellite feature consistent with low relative intensity of 0.05 estimated in this satellite for the 12W sample here (see Fig. S7 in Supplementary Material).

The resulting components appear with binding energies of Mo $3d_{5/2}$ peaks of H and T′ components at 228.6 and 229.3 eV in 0W sample (Fig. 3a) and at 228.4 and 229.1 eV in 12W sample (Fig. 3b). It is clear that T′ is the dominant phase at the selected pixel for both 0W and 12W samples, with contributions of ~ 75% and ~ 67%, respectively, of the total Mo $3d$ signal. It is to be noted that this decrease of the T′ contribution in the 12W sample compared to that in the 0W sample refers to the spots identified to have the highest T′ contributions on respective samples. The overall decrease in the T′ contribution in the 12W sample is evident from the map of the relative T′ intensity for the two samples presented in Fig. 2d and 2e. Therefore, a significant amount of T′ phase appears to convert back to the stable H phase on washing with water and the consequent removal of Li$^+$ ions.

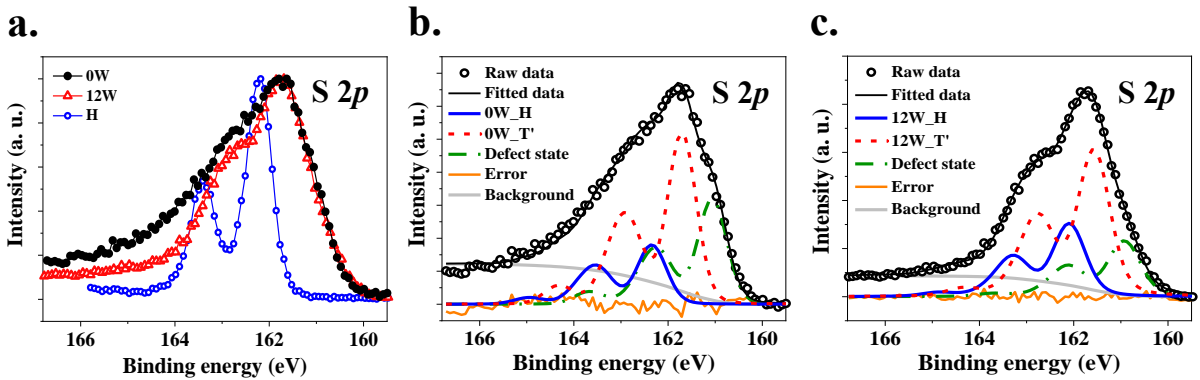

**Fig. 4.** (a) S $2p$ core level spectra for pristine H, 0W and 12W samples; Fitting of these spectra with three contributions for (b) 0W and (c) 12W samples.

We show the S 2$p$ spectral region from 0W (black dots), 12W (red triangles), and H (blue circles) samples in Fig. 4a. While the S 2$p$ spectrum of the H phase shows the expected spin-orbit split doublet structure, consisting of 2$p_{3/2}$ and 2$p_{1/2}$ signals with sharp features, the S spectra from the Li-treated 0W and 12W samples are rather broad and featureless, indicating the presence of multiple species of S$^{2-}$ ions in these samples. Substantial additional intensities on the lower binding energy side for the chemically treated samples compared to the H sample suggest the existence of the metastable T′ state as in the case of Mo 3$d$ spectral region (see Fig. 2a). Additional spectral intensities on the higher binding energy side for the treated samples in Fig. 4a, with more enhanced intensity for the 0W sample, is most likely due to the presence of prominent contributions from the plasmonic loss satellites, as also observed in the Mo 3$d$ spectral region. These close similarities prompt us to analyse the S 2$p$ spectral region in a way similar to that for the Mo 3$d$ region for the chemically treated samples, allowing for simultaneous presence of signals from two polymorphs, H and T′, along with plasmonic loss satellites accompanying main peaks with a constant intensity ratio and peak separation. Such an analysis, however, fails to provide a satisfactory description of the S 2$p$ spectral region, as shown in Fig. S8 (in Supplementary Material) suggesting the presence of other species. In order to keep the description as simple as possible, we allowed for an additional species of S$^{2-}$ to contribute in this spectral region. We find that the description in terms of contributions from three species of S$^{2-}$, namely originating from the H, the T′ and the additional S species, is sufficient to provide a consistent and convincing description of the entire spectral regions for both 0W and 12W samples. We show the outcome of this fitting procedure, resulting in the spectral decomposition for the S 2$p$ regions of 0W and 12W samples in Fig. 4b and 4c. The three spectral components are marked as H (blue solid line), T′ (red dashed line), and the additional, third component (green dash-dotted line) termed as the "defect". While the origin of H and T′ related signals are obvious, being related to patches of the chemically treated samples existing in the corresponding crystallographic forms, the appearance of the third component in the sulphur spectrum can be understood in the following manner. It is clear from our results that there is extensive coexistence of the H and T′ phases in such chemically treated samples. All T-derived phases, such as T, T′, T″ and T‴ phases, are structurally related to the H phase by a sliding arrangement of the top $S^{2-}$ layer to change the ABA arrangement of the hexagonal planes in the H phase to the ABC type of arrangement of the layers in all T derived polymorphs. Extensive coexistence of H and T′ phases would then suggest small patches (< 120 nm of the photon beam size) of these two polymorphic forms, necessitating extensive coexistence of ABA and ABC type arrangements in any given flake of chemically treated $MoS_2$. Thus, the top $S^{2-}$ layer would locally shift between the A type and the C type arrangements with respect to the bottom $S^{2-}$ layer in nanoscopic domains. The domain boundaries will represent defect lines, necessarily strained to accommodate the local sliding arrangement of the $S^{2-}$ layer. The $S^{2-}$ ions along this domain boundaries represent a "defect" $S^{2-}$ ions with a distinct electronic structure compared to $S^{2-}$ ions within the H-type or the T′-type polymorphic forms.

We note that S 2p region can be fitted reasonably well, considering the plasmons associated with it, in a way very similar to that of the Mo 3d fitting with relative intensities of 0.23 and 0.05 and relative peak positions of 1.5 and 1.2 eV with respect to the corresponding main peaks for the 0W and 12W samples, respectively. These values are almost same to that of the numbers obtained from Mo 3$d$ analysis, indicating the reliability of the fit parameters extracted here and the description of the spectral features in terms of the components. Thus, both Mo 3$d$ region and the S 2$p$ region suggest that the plasmonic satellite contributions decrease with the removal of the $Li^+$ ions via washing.

In order to understand the nature of changes in the electronic structure between 0W and 12W samples and the consequent enhanced stability of the T′ phase in the 0W sample, we carefully compare the components, obtained from the fitting of Mo 3$d$ spectra for both 0W and 12W samples in Fig. 5a by normalizing them at $3d_{5/2}$ peak; this procedure allows us to readily identify spectral shifts between different components as well as between different samples. Here solid and dotted lines refer to 0W and 12W samples, respectively, whereas H and T′ phases have been indicated using red and blue colour. Therefore, solid blue and red lines denote the Mo 3$d$ spectral positions from the T′ and H components of the 0W sample, while the dotted lines represent the corresponding components for the 12W sample. It is most interesting to note that the blue solid line and the dotted line, representing the T′ phase in 0W and 12W samples, respectively, show a shift of 0.2 eV for the dotted line of the 12W sample towards the lower binding energy compared to the blue solid line of the 0W sample. We note that this shift is evident even in the raw

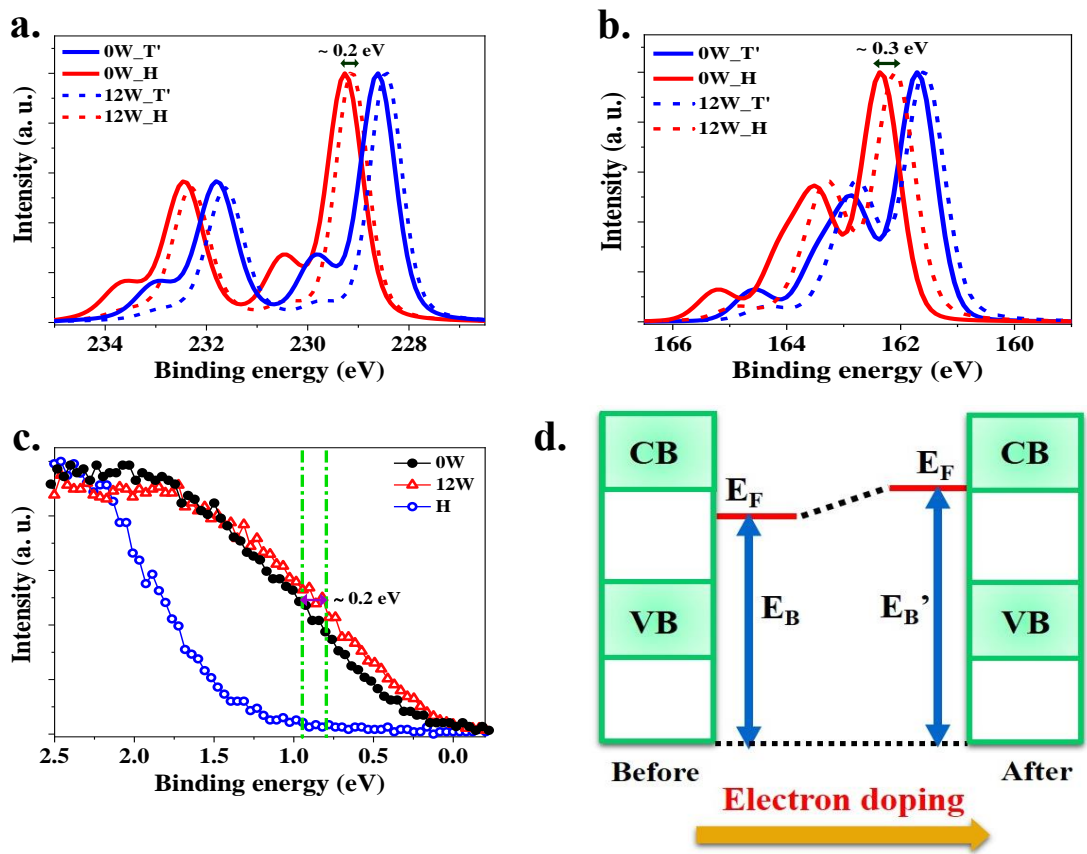

**Fig. 5.** H and T′ components of (a) Mo $3d$ core level spectra normalized at Mo $3d_{5/2}$ peak and (b) S $2p$ core level spectra normalized at S $2p_{3/2}$ peak for both 0W and 12W samples; (c) Spectra near the Fermi edges for pristine H, 0W and 12W sample; (d) Schematic diagram of the shift of the Fermi level due to electron doping of the system, reflecting as a shift in the binding energies of all levels.

experimental spectra of 0W and 12W samples, shown in Fig. 2a. We also note that the H components shown by the red solid line and the dotted line in Fig. 5a also show the identical 0.2 eV shift to the lower binding energy for the dotted line corresponding to the H component in the 12W sample. Similar plots are shown in Fig. 5b for S $2p$ components after normalizing them at the $2p_{3/2}$ peaks. Once again, we observe similar shifts of 0.3 eV towards lower binding energy side for all S $2p$ components of the 12W sample compared to those of 0W sample. Even the valence band edge of the 12W sample exhibits a 0.2 eV shift to the lower binding energy compared to that of the 0W sample, as illustrated in Fig. 5c. In Fig. 5c, we have also plotted the valence band edge region from the pristine H sample with blue circles, whereas the black dots and red triangles are used to plot the valence band edge of 0W and 12W samples, respectively. All three spectra shown in Fig. 5c exhibit negligible intensity at the Fermi energy ($E_F$), clearly suggesting that none of these samples can be regarded as a metal, for example as would be expected from the undistorted

T phase with a large, finite density of states at the $E_F$. Fig. 5c also shows that for pristine H sample the intensity is negligible down to ~ 1.0 eV binding energy, indicating a sizeable bandgap which in turn leads to the conclusion that the spectra from the intercalated samples within the range of binding energies smaller than ~ 1.0 eV are contributed only by the metastable, small bandgap T′ phase. The comparison of the spectral features clearly establishes that 0W sample has the leading edge of the valence band shifted towards the higher binding energy compared to that of the washed 12W sample by the energy separation comparable to that of the core level spectra.

This consistent spectral shift of about 0.2 eV towards the higher binding energy for all spectral features from the 0W sample cannot be explained by any mechanism other than in terms of some electron doping of the 0W sample, thereby moving up the Fermi energy. Since all spectra are referenced to the Fermi energy in photoelectron spectroscopy, an upward shift of this reference energy, as would happen with electron doping, will appear as a constant shift of the same magnitude towards the higher binding energy for all spectra, as shown schematically in Fig. 5d. The electron doping is most likely dominated by the presence of $Li^+$ ions, each ion donating one electron to $MoS_2$ and therefore, determining the pinning of the Fermi energy to some level. With the removal of Li via washing also removes the doped electrons, thereby moving the Fermi energy down, this phenomenon reflecting itself as a shift of all spectroscopic features to a lower binding energy, as observed. We may also relate this electron doping via $Li^+$ ions to the presence of prominent plasmon features only in the 0W sample. Clearly, the spectral features of the pure H sample show no signature of plasmonic loss satellites. Thoroughly washed 12W sample with $Li^+$ ions below the detection limit shows an evidence of only a very low intensity plasmonic replica of all spectral features. Thus, it appears that the additionally doped electrons via the presence of $Li^+$ ions on $MoS_2$, residing close to the bottom of the conduction band, contribute to the plasmonic loss features, observed prominently for the 0W sample. It is then easy to visualise that such electrons, doped via $Li^+$ ions, pinning the Fermi energy near the bottom of the conduction band will also make chemically exfoliated $MoS_2$ more conducting. This conductivity will depend on the extent of $Li^+$ ion induced electron doping, but is also aided by the formation of the small bandgap ($90 \pm 40$ meV) metastable T′ phase, such that the thermally excited free charge carriers from the Fermi energy to the conduction band will be significant in number. This scenario also suggests a possible explanation for our observation that the presence of $Li^+$ ions is essential for the extensive presence of the metastable state, as evidenced by a significant depletion of the metastable state on removal of Li$^+$ ions via washing (see Fig. 1a). It is obvious that electron doping will prefer a small bandgap semiconductor like the T′ phase, rather than one like the H phase with a considerably higher energy conduction band. Thus, the presence of Li$^+$ ions and the consequent electron doping of MoS$_2$ tilts the energy balance slightly in favour of the doped T′ phase, this advantage being absent in the undoped case.

The ready and irreversible conversion of the T′ phase to the stable H phase on various mild perturbations, such as warming the sample at around ~ 98°C [54], establishes that this phase is indeed a higher energy, or a metastable phase and the H phase is thermodynamically the ground state. So, the ability to reach this alternate state must be kinetically controlled. The fact that mechanical exfoliation does not yield this metastable phase and that the presence of Li$^+$ ions is necessary to reach this state suggests that Li$^+$ ions must be contributing to the kinetic pathways to reach the metastable state. After having reached the metastable state, the removal of Li$^+$ ions by washing leads to the partial conversion of the metastable state to the stable H phase. Since the metastable state was already kinetically stabilised during the Li$^+$ intercalation, the conversion of the metastable phase to the stable phase must be related to the change in the potential energy landscape with the removal of the Li$^+$ ions, possibly lowering the barrier locally for the conversion to the stable phase at the same temperature. We also argue that the presence of a doped electron in presence of Li$^+$ intercalation reduces the energy difference between the stable H phase with its large bandgap and the metastable T′ phase with its small bandgap in favour of the metastable phase, providing a plausible thermodynamic reason for the change in the local potential landscape between the Li$^+$ ion doped state and the Li$^+$ ion removed states.

4. Conclusion

In conclusion, we have shown that chemically exfoliated MoS$_2$ has an extensive presence of a distorted metastable T′ state that has a small bandgap compared to the stable H phase. Raman spectra clearly establish that this metastable state cannot be the undistorted T form, while the electron spectroscopy directly shows the absence of any significant metallic density of states at the Fermi energy. Investigating samples with different extent of washing to remove Li$^+$ ions, we show that Li$^+$ ions dope the small bandgap semiconducting metastable state to make these more conducting than in their pure state, explaining why such chemically exfoliated $MoS_2$ has often been termed as metallic and this in turn has led to the erroneous conclusion of the undistorted T phase formation. Our results also show that these doped electrons facilitate the formation of the metastable, small bandgap, distorted T′ phase, enhancing their relative stability via doping, this is at the origin of the extensive formation of this metastable state via the chemical exfoliation route.

Acknowledgement:

The authors thank Smruti Rekha Mahapatra for her help in AFM measurements and G V Sai Manohar for initial Raman characterization (not included in this manuscript). The authors thank the Department of Science and Technology, Government of India, Indo-Italian Program of Cooperation, and International Centre for Theoretical Physics, Trieste, for supporting this research. D.P. acknowledges the Council of Scientific and Industrial Research for a student fellowship. D.D.S. thanks Jamsetji Tata Trust for support.

# Supplementary Material

On the origin of metallicity and stability of the metastable phase in chemically exfoliated MoS$_2$

Debasmita Pariari [a], Rahul Mahavir Varma [a], Maya N. Nair [a,+], Patrick Zeller [b], Matteo Amati [b], Luca Gregoratti [b], Karuna Kar Nanda [c], D. D. Sarma [a,*]

[a]*Solid State and Structural Chemistry Unit, Indian Institute of Science, Bengaluru 560012, India*

[b]*Elettra-Sincrotrone Trieste S.C.p.A., SS14, km 163.5 in AREA Science Park, 34149 Basovizza, Trieste, Italy.*

[c]*Materials Research Centre, Indian Institute of Science, Bengaluru 560012, India*

[*]Corresponding author: sarma@iisc.ac.in

[+]Present address: CUNY Advanced Science Research Center, 85 St. Nicholas Terrace, New York, NY 10031, USA

1. **Structure of undistorted (T) and distorted (T') octahedral phase:**

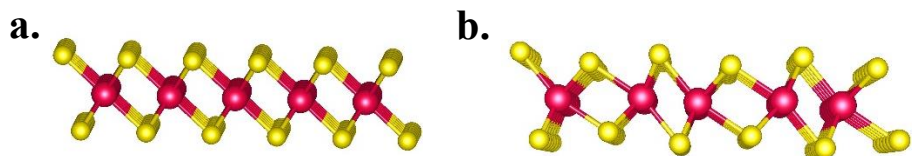

**Fig. S1.** Side views of (a) T and (b) T' phase of MoS$_2$ where red and yellow balls indicate Mo and S atoms, respectively.

Six S atoms are arranged around a Mo atom in a pure and distorted octahedral fashion in T and T' phases, respectively. For the T phase, Mo-Mo distance is same throughout the chain in contrast to the T' phase, where dimerization of consecutive Mo atoms leads to the formation of zigzag arrangements of the Mo ions. The side views for both T and T' phases are shown in Fig. S1a and S1b. These structural changes, representing a Jahn-Teller distortion of the $d^2$ configuration of Mo$^{4+}$ from a regular octahedron to a distorted one, changes the ground state electronic structure of MoS$_2$ from being metallic in T phase to a small bandgap semiconductor in the T' phase. Moreover, total energy calculations suggest that the distorted phase is relatively more stable compared to the undistorted state; in fact, the undistorted state is found to be unstable and is expected to spontaneously distort.

## 2. AFM study on the pristine and intercalated flakes:

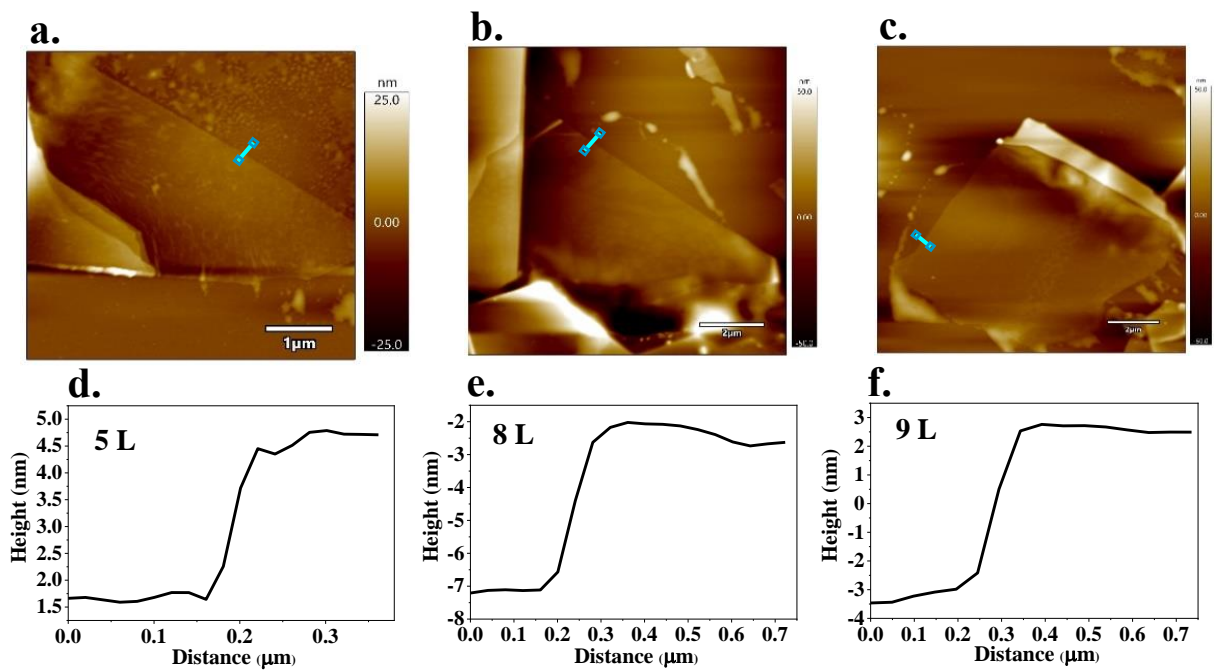

**Fig. S2.** (a), (b), (c) Representative AFM images of the pristine H samples and (d), (e), (f) corresponding height profiles of the flakes represented in (a), (b) and (c) respectively through the cyan lines.

Mechanically exfoliated H samples (Fig. S2) were found primarily in the range of 4-9 layers with characteristic layer separation of about 0.65 nm, as reported in the literature [1,2].

In order to illustrate the effect of $Li^+$ ion intercalation as well as washing these intercalated flakes, we show in Fig. S3 an individual flake of pristine H (also shown in Fig. S2c) sample before intercalation (Fig. S3a), after $Li^+$ intercalation (Fig. S3b) and after washing with water (Fig. S3c). It clearly shows that the interlayer separation is increased to nearly 1.7 nm on intercalation, which reduces somewhat to about 1.1 nm on substantial removal of $Li^+$ ions via washing, as has been reported in the literature [3]. These hugely enhanced interlayer separation decouples successive layers, leading to essentially non-interacting single layers of $MoS_2$ in these chemically treated samples.

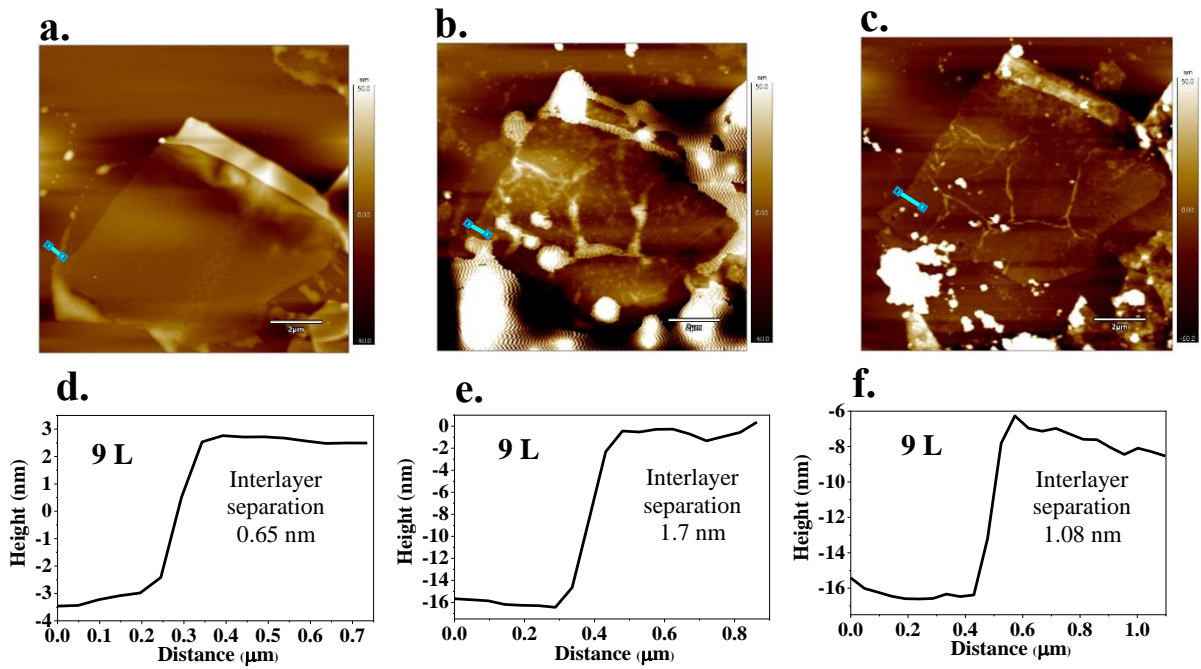

**Fig. S3.** Representative AFM images of an individual flake (a) before intercalation (H), (b) after Li$^+$ intercalation (0W) and (c) after washing (12W) with corresponding height profiles represented by the cyan lines shown in (d), (e) and (f) respectively.

## 3. Evidence for the presence of lithium ions (Li$^+$):

In Fig. S4, photoelectron spectra in the Li 1$s$ region for the four chemically exfoliated samples, namely 0W, 1W, 6W and 12W, are presented along with the spectrum from the mechanically exfoliated H sample. The 0W sample exhibits one peak, responsible for Li 1$s$, at binding energy of ~ 55.8 eV, shown in Fig. S4a normalized at the peak position. Two representative data, indicated by black and red lines, taken on different samples, are shown here in order to prove the reproducibility of the data. The spectrum in the binding energy window responsible for Li 1$s$, obtained from other chemically exfoliated samples apart from 0W and pristine H sample, are compared in Fig. S4b. It is quite clear that, only the noise level can be visible in Li 1$s$ binding energy window for these samples, which further proves that the intercalated Li$^+$ ions react with water so vigorously that, even after a single wash the concentration level of Li$^+$ goes beyond the detection limit.

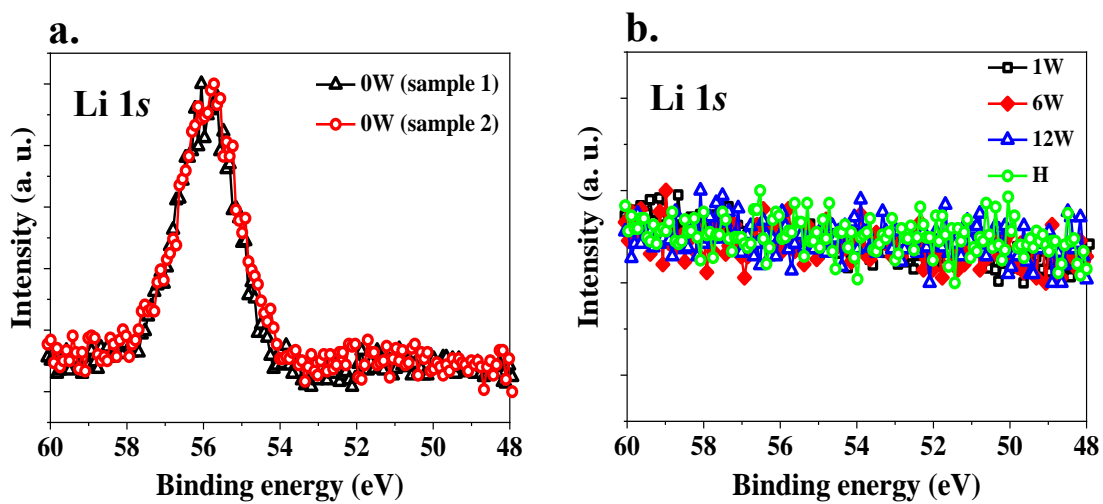

**Fig. S4.** Li 1$s$ core level spectra for (a) 0W samples and (b) 1W, 6W and 12W and mechanically exfoliated pristine H samples.

### 4. Raman spectra of the samples with different washing cycles:

We carried out preliminary studies for several control groups, namely 1W, 4W and 6W, in addition to 0W and 12W samples. The similarity of all washed samples is ascertained by the Raman spectra, (shown in Fig. S5).

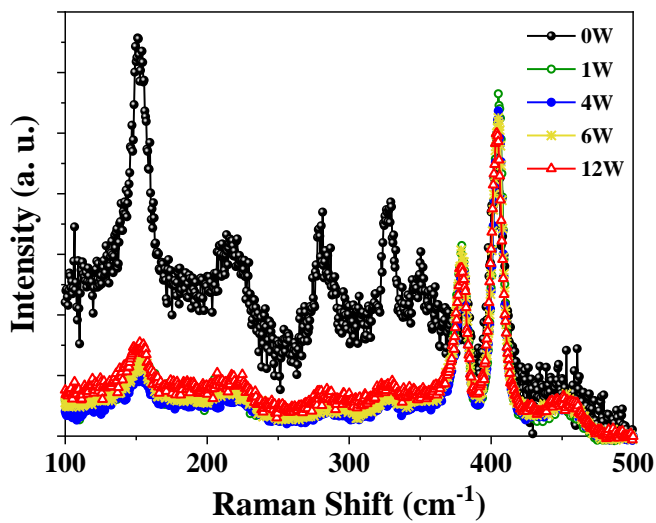

**Fig. S5.** Comparison of room temperature Raman spectra for the samples with different washing cycles normalized at 405 cm$^{-1}$ (A$_{1g}$ peak).

## 5. Photoelectron spectra of the samples with different washing cycles:

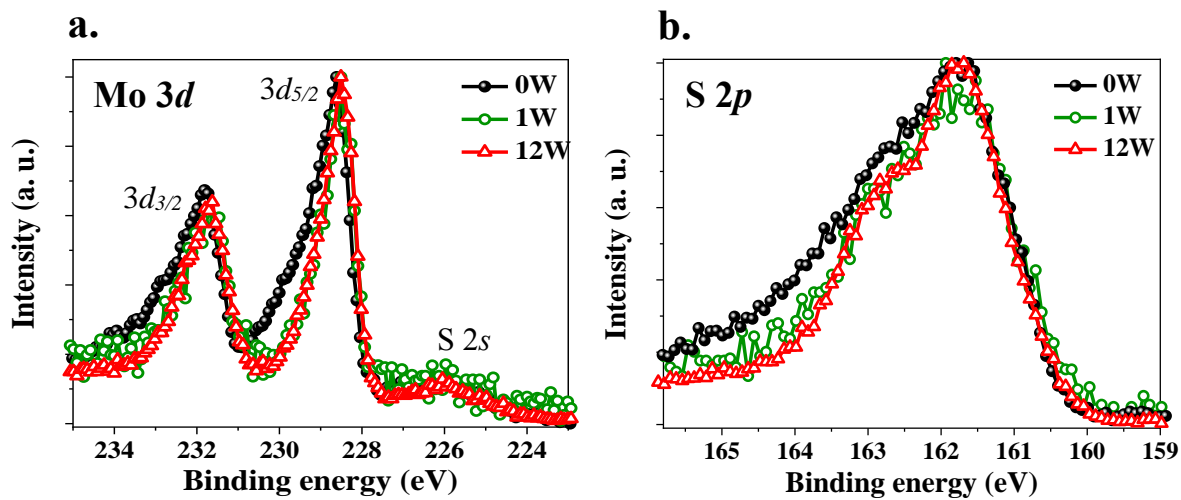

**Fig. S6.** Comparison of photoelectron spectra for 0W, 1W and 12W samples in the energy range of (a) Mo 3*d* and (b) S 2*p* core level.

The similarity of the two limiting washed samples, namely 1W and 12W, are distinct from the 0W sample in terms of their photoelectron spectroscopic results as shown in Fig. S6a and S6b.

## 6. Analysis of Raman spectrum from Group Theory point of view:

The presence of extra Raman peaks in intercalated samples can be explained with the help of Group Theory. The point groups, describing the symmetry of single MoS₂ layer of pristine H, T and T' phases, are *D₃ₕ*, *D₃d* and *C₂ₕ* respectively. For each of the above mentioned point groups, irreducible representations corresponding to the normal modes of vibrations at zone centre (k = 0) along with the Raman and IR active modes are given below:

**Table 1:** Point group symmetry for different phases of MoS₂ along with the comparison between number of Raman active modes predicted from Group Theory and experimental observations.

| Point group | Irreducible representation | IR active | Raman active | No. of Raman modes | No. of prominent Raman peaks observed experimentally |
|---|---|---|---|---|---|
| D₃ₕ | $\Gamma_{vib}(\text{H}) = A_1' + 2E' + 2A_2'' + E''$ | $E', A_2''$ | $A_1', E', E''$ | 3 | Mostly 2. Presence of very weak $E''$ mode i.e. $E_{1g}$ peak is reported in few studies[4,5]. |
| D₃d | $\Gamma_{vib}(\text{T}) = A_{1g} + E_g + 2E_u + 2A_{2u}$ | $E_u, A_{2u}$ | $A_{1g}, E_g$ | 2 | 2 peaks[5] |
| C₂ₕ | $\Gamma_{vib}(\text{T}') = 6A_g + 3B_g + 3A_u + 6B_u$ | $A_u, B_u$ | $A_g, B_g$ | 9 | 6 peaks*[6,7] |

\* Till now as per our knowledge there is no experimental report of stabilizing pure T' phase. It is always coexisted with stable H phase. Therefore, it is inevitable to get signature of H phase in Raman spectroscopy, from the samples containing T' phase. For such samples, three peaks responsible for H phase are accompanied by three more peaks, presumably coming from the presence of T' phase.

## 7. Spectral fitting of 12W sample without plasmonic peak:

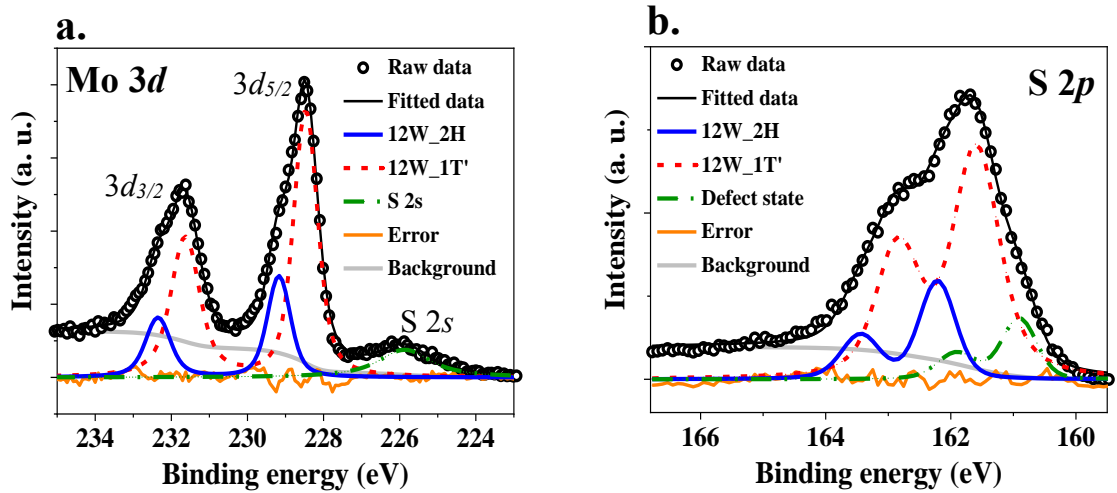

**Fig. S7.** Fitting of (a) Mo 3*d* and (b) S 2*p* core level spectra for 12W sample without considering the plasmon loss features.

Both Mo 3*d* and S 2*p* spectra obtained from 12W sample can be fitted with good agreement without considering the plasmon loss features, showing in Fig. S7. This clearly indicates towards almost non-existent nature of the plasmon satellites in 12W sample.

## 8. Fitting the core level spectra:

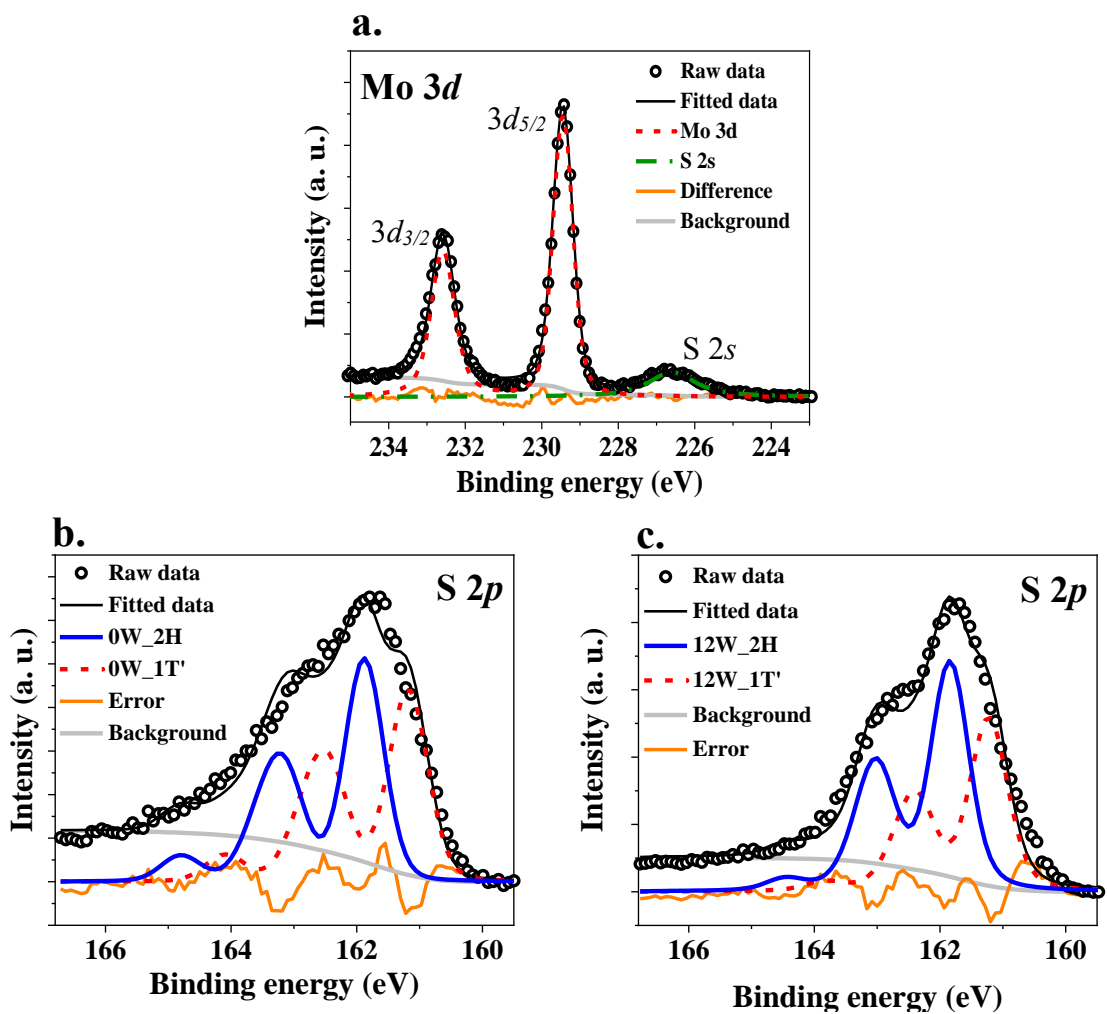

**Fig. S8.** (a) Fitted Mo 3$d$ core level spectra obtained from mechanically exfoliated sample; S 2$p$ core level spectra have been fitted with two components for (b) 0W and (c) 12W samples.

As there is only H phase present in mechanically exfoliated sample, the spectra can be decomposed in Mo 3$d_{5/2}$, Mo 3$d_{3/2}$ and S 2$s$ component, showing in Fig. S8a. The shape of each peak is defined by a Lorentzian function convoluted by a Gaussian function, representing the lifetime and resolution broadening respectively. Throughout our analysis we have used a single Gaussian broadening for all components as the source of the resolution broadening lies in the instrumental artefact. We have imposed the constrain of fixed relative intensity between each spin orbit coupling pair i.e. the intensity ratio of 3$d_{5/2}$ to 3$d_{3/2}$ is 1.5 and similarly the intensity ratio between 2$p_{3/2}$ and 2$p_{1/2}$ is 2, along with the constant energy separation between each of these spin orbit pairs. We have used the least number of components required for a reasonable description of the Mo 3$d$ and

S 2*p* spectra obtained from chemically exfoliated samples. As shown in Fig. S8a, only one component is sufficient to fit the spectra, while the Mo 3*d* spectra for both 0W and 12W samples can be fitted with the help of only two components namely, H and T' phase. It appears to be impossible to fit S 2*p* spectra assuming only two components as shown in Fig. S8b and S8c. For both 0W and 12W samples, the relative abundance of the H component is higher compared to that of the T' component, which is contradicting the results obtained from Mo 3*d* fitting. The spin orbit coupling components in Mo 3*d* spectra are clearly distinguishable, indicating a much reliable fit in that case. On the contrary, 2*p*$_{3/2}$ and 2*p*$_{1/2}$ peaks are merged together for both the intercalated samples, making the fitting more difficult. In this scenario it is obvious to stick to the relative intensity value of the two components obtained from Mo 3*d* fitting, which in turn makes the S 2*p* spectral fitting with two components obsolete.

As described in the main text, in order to take care of the extra intensity, appearing in the higher binding energy side of each peaks in Mo 3*d* spectra of the 0W sample, we have assumed the presence of plasmon satellites. That can be easily done by defining a function, which is the summation of four Lorentzian functions. It is important to note that we have kept the intensity ratio fixed between two spin orbit coupling pairs of both the main and the plasmon peaks but the relative intensity between a main peak and its plasmon has been allowed to vary. Similarly, the peak positions have been varied in such a way that the distance from the main peak to its plasmon for both the spin orbit coupling components remain same.